\documentclass[a4paper, 11pt]{article} 

\usepackage[protrusion=true,expansion=true]{microtype} 
\usepackage[english]{babel}
\usepackage{graphicx} 
\usepackage{wrapfig} 
\usepackage{amsmath,amsthm,amssymb}
\usepackage{hyperref}
\usepackage{natbib}
\usepackage[textwidth=90]{todonotes}
\usepackage{mathpazo} 
\usepackage[utf8]{inputenc} 
\usepackage[footnotesize]{caption}
\usepackage{subcaption}
\usepackage{float}
\usepackage{csvsimple}
\usepackage{booktabs} 
\usepackage{siunitx} 
\usepackage{pgfplotstable} 
\usepackage{enumerate}

\makeatletter
\renewcommand\@biblabel[1]{\textbf{#1.}} 
\renewcommand{\@listI}{\itemsep=0pt} 

\renewcommand{\maketitle}{ 
\begin{flushright} 
{\LARGE\@title} 

\vspace{50pt} 

{\large\@author} 
\\\@date 

\vspace{40pt} 
\end{flushright}
}

\pgfplotsset{compat=1.12}

\title{\textbf{An Agent-based Model of Contagion in Financial Networks}} 

\author{\textsc{Leonardo dos Santos Pinheiro} 
	\\{\textsc{Flávio Codeço Coelho}}
\\{\textit{Getulio Vargas Foundation}}} 

\date{\today} 

\begin{document}

\maketitle 



\begin{abstract}
This work develops an agent-based model for the study of how the leverage through the use of repurchase agreements can function as a mechanism for the propagation and amplification of financial shocks in a financial system. Based on the analysis of financial intermediaries in the repo and interbank lending markets during the 2007-08 financial crisis we develop a model that can be used to simulate the dynamics of financial contagion.
\end{abstract}

\hspace*{3,6mm}\textit{Keywords:} agent-based models, financial risk, computational economics, financial contagion 

\vspace{30pt} 


\section{Introduction}


In recent years, the use of complex models for the analysis of financial contagion  in economic systems has become widely used. The recent 2007-08 financial crisis, regularly attributed to the complex  relationships among financial institutions, has revived the interest in the role of the behavior of market participants  in the creation of the complex interlinkages which serve as channels for the transmission and amplification of economic shocks.

The  crisis, which started with a liquidity drain in the US sub-prime mortgage market, due to the collapse of a bubble in the housing market, quickly overflowed to debt markets and stock markets in a process of financial contagion that eventually prompted the downfall of major American and European banks and triggered a world recession. The means by which the crisis spread from a specific bubble to the whole financial system is what we call financial contagion and it is a process made possible by the existing interconnectivity between financial institutions.

Financial institutions are interconnected in a variety of ways, both directly and indirectly. Direct interconnectedness happen mostly through mutual credit exposures while indirect interconnectedness occurs mainly through common asset holdings, margin call losses and haircut increases triggered by fire sales and liquidity drain and information spillover (\citet{liu2015banking}).

Direct interconnectedness occurs because mutual credit exposures between financial institutions can lead to domino effects. With the complex chains of intermediation which exist in the global financial system, the failure of a highly interconnected institution can cause major disruptions to the financial system as a whole as this institution wouldn't be able to fulfill it's obligations and cause mark-to-market losses in the balance sheets of all other institutions with direct exposure to it, which could cause a number of other institutions to face distress as well.

Indirect interconnectedness occurs as institutions facing distress can start fire selling it's assets. Fire sales further stress the market prices of the assets owned by the company, causing mark-to-market losses in all institutions with common asset holdings and causing increases in margin calls and haircuts in repurchase agreements backed by these assets. Information spillover can also cause other institutions with similar balance sheets to face higher spreads.

As more institutions suffer losses and become distressed, market conditions may further deteriorate via the aforementioned contagion channels, leading to a negative feedback loop and, possibly, to a cascade of failures. 

While many approches to understand the dynamics of financial contagion using equation-based modeling have been developed, mostly through economic and network models (see \citet{gai2010contagion}, \citet{huang2013cascading} and \citet{elliott2014financial}), these approaches have the limitation of reproducing an homogenized and simplified approximation of the observed reality, sometimes producing unrealistic models which are not sufficiently justified (\citet{helbing2010fundamental}). 

In this work we focus on the prospects of the computer simulation of economic systems to model the dynamics of financial contagion. Agent-based modeling is a computational technique where the components of a system are encapsulated as agents, which can represent individuals, groups, companies and/or countries, while  the analysis of the system is carried out through the interactions of these agents (\citet{helbing2012social}).

By modeling the financial system through the use of agents, we are capable not only of  creating simulations that reflect the interactions between different entities more accurately, but also of testing the implications of different hypothesis. We furthermore emphasize the importance of building models using a range of empirical observations to design more realistic models which  are capable of representing market dynamics observed in historical episodes and allows us to explore in more detail the dynamics of financial markets. 

In this work we focus on modeling one of the most prominent effects of the 2007-08 financial crisis: the liquidity drain observed in the repurchase agreement (repo) markets. During the crisis both interbank lending and repurchase agreements shrank dramatically, causing a massive deleverage in the financial system and threatening several banks with insolvency in a movement that only stopped through a government bailout. To achieve this objective we focus on the interaction between banks, money market funds and hedge funds in the repo and interbank markets in order to recreate this financial contagion movement.

The remainder of this paper is structured as follows. In Section \ref{sec:fincont} we introduce the problem of financial contagion and focus on the repo markets. Section \ref{sec:abmodel} discusses agent-based models of financial markets and how they can be used to understand market dynamics such as the one we wish to model. Section \ref{sec:model} presents our agent-based financial contagion model. In Section \ref{sec:simulation} we perform some numerical simulations and discuss the results. Finally, in Section \ref{sec:conclusion} we discuss extensions of the proposed method and our conclusions.

\section{Financial Contagion}\label{sec:fincont}

Strong financial contagion has been one of the key features of most recent financial crises, as localized problems in certain segments of the markets spread to other segments leading to the risk of cascading defaults and failures which are often avoided through government bailouts of institutions deemed "too big to fail."

As described by \citet{gorton2009run}, the panic of 2007-08 occurred through a run on the repo market. The repo market is very important market that provides collaterized financing for banks.They work very much like bank deposits, but for firms operating in the capital markets. In a repurchase agreement the bank sells a security  with the promise of repurchasing the security at a specified price in the end of the contract. The intermediary buying the security from the bank is remunerated by the spread in operation.

According to \citet{gorton2009run}, in the last twenty-five years a number of financial innovations have allowed traditional assets of banks to be traded in capital markets through securitization and loan sales and have allowed banks to leverage through these operations.

Since the 2007-08 crisis, the interconnected nature of financial markets has not only been studied as an explanation for the spread of risk and losses throughout the system, but also motivated much of the policy recommendations in the aftermath.  Yet, a framework to understand how the dynamics of the network structure of the financial market, specially the repo market, leads to systemic risk remains incomplete.

In a broader sense, there is currently a high level of uncertainty about which elements in the structure of the financial system causes contagion and how it occurs. Early work, prior to the crisis, focused on general aspects of interbank lending such as the work of \citet{allen2000financial}, which modeled contagion as an equilibrium phenomenon caused by liquidity preference shocks through economic regions, and of \citet{rochet1996interbank}, which considers the systemic risk created by interbank lending and investigates whether decentralized bank interactions can be preserved while maintaining the stability of the system.

More recent work, such as \citet{gai2010contagion}, \citet{acemoglu2013systemic} and \citet{elliott2014financial} examine how shocks propagate through a network based on debt holdings or interbank lending and, also, how shocks propagate as a function of network architecture.


While these works have provided useful insights about financial contagion (although presenting quite different and complementary results), the use of economic equilibrium and network models have some limitations in the study of the phenomenon. For instance, financial agents usually have different goals and strategies, thus, behaving very differently. Also, we must consider that the nature of debt exposures as connectivity measures can also vary greatly, with mutual lending exposures, cross-holding of shares, repurchase agreements and common asset holdinds of other sorts (e.g. stocks) having a different impact on the propagation of shocks.

Accounting for these heterogeneities in  network and economic models can lead to mathematical intractability very fast. A different approach, which may lead to a more accurate representation of the financial system, despite being unable to render an analytical solution to understand the problem, is to use agent-based simulation, as we describe bellow.

\section{Agent-Based Computational Finance} \label{sec:abmodel}

Much of the work in economics and finance hopes to simplify human interactions and behaviors in a way that we can analyze these systems through aggregated macro-features. But complex systems involves complex interactions among many individuals and, in some cases, this complexity makes the use of analytical models to understand the system unfeasible. For this reason, agent-based models and simulations have become an invaluable tool for understanding the dynamics of the economic and/or financial system as a whole.

Agent-based models are a class of computational models used to simulate the actions and interactions of autonomous agents (\citet{gilbert2008agent}). In computational economics, these models have been used to study properties of markets by building and simulating markets, especially in the field of computational finance and there are many ways in which agent-based models can be used to study financial markets \footnote{A broad discussion of how market microstructure can be studied by simulations with intelligent agents can be found in \citet{chan2001artificial},  while \citet{lebaron2006agent}  discusses how agent-based models can be used to build artificial markets  and \citet{tsang2004computational} discusses other examples of agent-based models uses in computational finance}

The building of artificial markets is one of the most important contributions of agent-based models to the study of financial systems. They allow us to model economic agents according to a theoretical model and to observe if our economic assumptions about the agents interactions in a financial setting would generate the expected dynamics. 

Since the eighties some models of artificial markets have been tried, specially for stock markets. \citet{cohen1983friction} tried to look the impact of random behaving agents on various market structures, while \citet{kim1989investment} used discrete event simulation to model the interactions of different kinds of trading agents and \citet{de1995exchange}  focused on the dynamics of foreign exchange markets.

One of most notable and most sophisticated markets is the Santa Fe Institute (SFI) market.  The SFI market was created with the idea of modeling a financial market with an ecology of trading strategies (\citet{lebaron2002building}).  The SFI Market structure were modeled to consider preferences and risk aversion in trading and even allowed the emergence of trading patterns over time through the use of genetic algorithms. Although there have been several generations of the SFI artificial market, consisting of modifications of the  market structure and of different programming platforms, the fundamentals of the theoretical model have persisted \footnote{A description of the SFI market model can be found in \citet{arthur1996asset} and \citet{palmer1994artificial} while an overview of the SFI market history can be found in \citet{lebaron2002building}) and in \citet{johnson2002agent}}.

Other artificial stock markets have been designed focusing on features not included in the SFI artificial market model. For instance, \citet{lebaron2001evolution} and \citet{lebaron2001empirical} have used a new framework including varying forecasting horizons and memory lengths, which is crucial in the convergence to a rational expectations equilibrium, while \citet{serguieva2007computational} have investigated herding behaviors as a possible reason for contagion among different markets, and  \citet{martinez2009heterogeneous} have elaborated an artificial market in which trading behaviors model technical, fundamental and noise traders, being able to recreate statistical properties of price series in real financial markets.

Outside of stock markets, \citet{arciero2008exploring} developed a model of real time gross settlement paying system for predicting the impact of disruptive events in the flow of interbank payments and \citet{llacay2010realistic} developed a model to simulate crisis and risk management in fixed-income markets.

Agent-based models of financial markets have allows to simulate and recreate episodes observed in historical data to assess economic theories. In this work we focus on building an artificial repo market and it's behavior under a liquidity shock.

\section{The Artificial Repo Market Model} \label{sec:model}

To simulate the dynamics of financial contagion in the repo market we build an artificial financial market where financial agents must manage their risk and may face defaults and bankruptcy if there are significant imbalances between their balance sheets. The financial risk is measured and controlled trough losses, liquidity and leverage metrics. 


Our artificial market structure is designed to reflect financial intermediaries that may choose to invest in a set of tradeable assets from outside the financial system (representing economic projects) and that can also make operations among themselves to improve resource allocation.

We design three types of financial intermediaries as agents, which can be banks, money market funds (MMFs) or hedge funds. These intermediaries interact with each other trading assets according to their roles, as described bellow, and with an optimization strategy. Every intermediary tries to maximize their gains while managing their risk.

\subsection{Assets}

For the assets that can be traded by the agents, there is a risk free government bond, a stock, representing a risky liquid asset, and a risky fixed-income asset (from this point only called risky asset), representing a economic project financed and securitized by banks. The intermediaries can also trade resources through interbank lending and repurchase agreements. These serve as instruments for them to improve resource allocation, and manage risk.

\noindent \textbf{List of Assets:}

\begin{enumerate}[A)]
	\item Government Bond
    
    In our market there is a government bond, consisting of a risk free asset, paying a constant interest rate, $r_f = 0.10$. This asset has complete liquidity as there is we assume there are external agents willing to match the order imbalance (treasury, foreign investors, central banks, etc.).
    
    \item Stock
    
    There is also a risky stock, similar to the one described in \citet{lebaron2002building}, paying stochastic dividend  following the autoregressive process: 
    
    \begin{align}\label{ref: eq_div}
		d_t = \bar{d} + \rho(d_{t-1} - \bar{d}) + \mu_t
	\end{align}
    with $\bar{d} = 10$, $\rho = 0.95$ and $\mu_t \sim N(0, \sigma^{2}_{\mu})$. The price of the stock is determined endogenously in the market.

    \item Risky Asset
    
    There is a risky asset paying a constant interest rate $r_r = 0.11$. This asset represents an economic project financed by banks and securitized in the capital markets. This asset can lose liquidity fast and may be a major source of risk. Since we do not implement mark-to-market calculation of bond prices, mainly because there are no variation in Government Bond interest rates, the higher interest rate reflects exclusively the perceived liquidity risk and the default risk of the asset.
    
	\item Interbank Loan
    
    Interbank lending play a key role in the financial system. They are vital for banks’ liquidity management.The interbank lending market is constituted by unsecured loans (the interbank loan) and secured loans (through repurchase agreements and described bellow). 
    
    The interbank loan is an operation where banks extend loans to one another for a small term. In our model they are used when banks don't have access to secured loans and must meet liquidity or cash requirements to avoid a default. The interbank loan has an interest rater $r_{IL} = r_{f} + \delta_i$ where $\delta_i$ is the risk premium paid by the borrower and:
    
\begin{align}
	\delta_i = \dfrac{\sum_j IL_{ij}}{\sum A_i \times P_{sell}(A)}
\end{align}

where $IL_{ij}$ is the value of interbank loan issued from bank $j$ to bank $i$, $A_i$ is the total value of asset $A$ owned by the bank $i$ and $P_{sell}(A_i)$ is the probability of selling the asset $A$ at each timestep, which is determined by the liquidity index of the asset, defined in Subsection \ref{ref:subsec_risk}. Interbank loans are always overnight.

    \item Repurchase Agreement
    
    Repos are a key mechanism in our fixed-income market. Repos require margining practices, where the borrower pays an initial margin, or ‘haircut’, to provide some protection to the lender in case the other party defaults.
    
    In our market, we implement a simplified version of repo operations\footnote{See \citet{livingston1999bonds} for a more complete explanation of repo operations in real markets.}. Repos can be backed up by Government Bonds or by the Risky Asset. Also, repos, as interbank loans, are always overnight, but can be renewed at each time step. We also implement margining pratices with the haircut being calculated as: 
    
\begin{align}
Haircut = 1 -P_{sell}
\end{align}
    
This means that assets with full liquidity will not need a haircut, while totally illiquid assets will require a 100\% haircut, meaning the asset is not acceptable as collateral. The repo is not renewed if the bank can not meet the margin call. If the repo is not renewed the bank must repay the repo, selling assets in the process if there is no available cash.

\end{enumerate}

\subsection{Financial Agents}

Each financial agent type has a special role in the model. Banks intact with the economic system by securitizing credit assets. Hedge funds mainly operate in the stock market to provide volatility. MMFs interact with banks in the money markets providing funding.

\noindent \textbf{List of Financial Agents:}

\begin{enumerate}[A)]
\item Banks

Banks are the central agents of our simulation. Banks can acquire the risky asset at will (it represents the interaction of the bank with the economic system) and may access the money market (composed by MMFs) to obtain funding. Banks may also lend directly to other banks, these operations are done in order to manage short-term liquidity.

\item Money Market Funds (MMF)

MMFs are mutual funds that serve as important intermediaries between investors who want highly liquid investments and banks that have short-term liquidity needs (\citet{rosengren2012money}). As described by \citet{rubin1999hedge}, MMFs are usually prevented by regulation from engaging in leveraging practices and investing in illiquid assets.

In our model, MMFs act buying short-term and liquid debt instruments, such as government securities, and providing liquidity to other market players through repurchase agreements.

\item Hedge Funds

In our model hedge funds act as minor players. They operate in the stock market and in the government bond market.
\end{enumerate}

\subsubsection{Trading strategies}

\noindent \textbf{Stock Trading}

The role of the stock market is to provide a major source of volatility and market risk to financial agents. \citet{schwert2011stock} have shown that stock markets present high levels of volatility following periods of market stress. Moreover, \citet{ben2012hedge} point out to the presence of severe funding constraints, such as investor withdrawals and lender pressure accounting, together with flight to quality (selloffs of high-volatility stocks) and static liquidity management (selloffs of high-liquidity stocks) as major sources of volatility and market downfall. These mechanisms play an important role the modeling of the financial contagion.

For stock trading, the abstract model we used to simulate a real market is similar to the Co-Evolutionary Heterogeneous Artificial Stock Market (CHASM) described by \citet{martinez2009heterogeneous}, but with some key differences in trader behavior, order types and market mechanism.

Stock trading is composed by classes of traders that are used in the literature and that represent trading behaviors in close resemblance to empirical observations. They help recreate the statistical properties of a real market:

\begin{enumerate}
\item \textit{Noise traders:} The noise trader represent a stock trader who lacks access to inside information and make erratic and irrational decisions (\citet{de1990noise}). 

A noise trader does not have any specific information about the security or is not capable of making adequate analysis of the information available to the market. If the efficient market hypothesis holds, noise traders add liquidity to a market by increasing the trading volume.

In our artificial market model, these traders buy, sell, or hold their positions with different probabilities $p_b$, $p_s$ and $p_h$, respectively. Each trader has it's own probabilities and the probabilities are defined before the simulations.

\item \textit{Fundamental traders:} Fundamental traders are stock traders which adhere to the principle of fundamental trading or value investing. They compare the true value they assess to the security with the current market price of the security. They seek to buy securities for which the market price is lower than the estimated true value (underpriced) and to sell assets for which the price is bigger than the estimated true value (overpriced).

The way fundamental traders are modeled is similar to the work of \citet{tsang2004computational}, they measure the true value of the stock by the dividend being paid by the security (See Eq. \ref{ref: eq_div}) and will change their position if the actual price deviates from the perceived true value by a threshold value of $\tau$. Each agent has a $\tau$ value drawn from an uniform interval $\left[ \tau_{\min}, \tau_{\max} \right]$

\item \textit{Technical traders:} Technical traders use technical and momentum indicators in the form of decision rules to forecast future stock prices. The list of indicators is taken from \citet{martinez2009heterogeneous} and are as follows:

\textit{Moving Average}:
\begin{align}
	MA(L, t) = \dfrac{P(t)- \left(\dfrac{1}{L} \sum_{i=1}^{L} P(t-i)\right )}{\dfrac{1}{L} \sum_{i=1}^{L} P(t-i)}
\end{align}

\textit{Trading Breakout}:
\begin{align}
	TRB(L, t) = \dfrac{P(t)- \max \{P(t - 1),...,P(t - L)\}}{\max \{P(t - 1),...,P(t - L)\}}
\end{align}

\textit{Filter}:
\begin{align}
	Filter(L, t) = \dfrac{P(t)- \min \{P(t - 1),...,P(t - L)\}}{\min \{P(t - 1),...,P(t - L)\}}
\end{align}

\textit{Volatility}:
\begin{align}
	Vol(L, t) = \dfrac{\sigma \{P(t - 1),...,P(t - L + 1)\}}{\dfrac{1}{L} \sum_{i=1}^{n} P(t-i)}
\end{align}

\textit{Momentum}:
\begin{align}
	Mom(L, t) = P(t) - P(t - L)
\end{align}

\textit{Moving Average Momentum}:
\begin{align}
	MomMA(L, t) = \dfrac{1}{L} \sum_{i=1}^{L} Mom(L, t - i)
\end{align}

The parameter L and and the forecasting rules are learned by Genetic Programming and the method is described in \citet{martinez2009heterogeneous}, \citet{tsang1998eddie} and \citet{li1999investment}.

\end{enumerate}

\noindent \textbf{Fixed-Income trading}

For fixed-income assets, while it is possible to use a sophisticated trading environment emulating bond arbitrage strategies, such as the one used by \citet{llacay2010realistic}, we opt for a more simpler and straightforward model.

All bonds (government bonds, risky asset) are traded in over-the-counter markets, that is, by trades directly between parties. For simplicity we assume that Government Bonds have complete liquidity and the order imbalance will be matched by market players outside the model.

The Risky Asset can be acquired at will by banks, but cannot be sold easily, they also have long maturity. Whenever a bank wishes to sell its position in the risky asset, it must find another financial agent willing to acquire it.

\noindent \textbf{Interbank lending and Repurchase agreement operations}

Secured and unsecured lending operations are carried out in a centralized manner. Banks use repurchase agreements as a leverage mechanism to acquire funding in the money market. Interbank loans are carried out whenever a bank is meeting short term liquidity issues.

\subsection{Risk Management}\label{ref:subsec_risk}

Agents in our model manage their daily market risk, liquidity risk and leverage to avoid bankruptcy.

\noindent \textbf{Market Risk}

Market risk is measured by Value-at-Risk (VaR). The VaR of a portfolio is the potential loss in value of the portfolio over a defined period for a given confidence interval. In general terms, the maximum loss is calculated as a function of the portfolio historical volatility $\sigma_p$ (\citet{borodovsky2000professional}). In the model all financial agents use the VaR Delta-Normal approach described in  \citet{linsmeier2000value}. In this approach, given that a financial agent will not tolerate a loss that will be equaled or exceeded by $x$ percent in a given time interval and the VaR is measured as:

\begin{align}
	VaR = - [ \mu - x \sigma]
\end{align}

where $\mu$ is the expected change in portfolio value. 

\noindent \textbf{Liquidity Risk}

The financial concept of liquidity is elusive. Market participants perceive a financial asset as liquid if they can sell a large amount of the asset without affecting its price, but this concept is hard to quantify and there is no single theoretically and universally accepted measure for liquidity (\citet{lybek2002measuring}).

As our liquidity metric we use the order imbalance to measure the probability of selling a asset at a given time step $t$ as:

\begin{align}
	P_{sell}(t) = 1 - \dfrac{\max \{0, (\sum_{i=i}^{n} \dfrac{O_S(t-i) - O_B(t-i)}{n})\}}{\sum_{i=i}^{n} \dfrac{O_S(t-i)}{n}}
\end{align}
where $O_S(t-i)$ is the volume of buy orders and $O_B(t-i)$ is the volume of sell orders at time step $i$ and $n$ is the horizon of calculation. This simple metric measures the probability that a agent wanting to sell an asset will find a buyer for the asset. The horizon for looking past orders must be short in order to capture the quickness of market liquidity deterioration.

\subsubsection{Leverage}

Leverage is one of the most critical sources of risk in our model. We use a simple leverage metric of: 

\begin{align}
	Leverage = \dfrac{TA}{NAV}
\end{align}

where $TA$ stands for total assets and $NAV$ is the net asset value of the agent.

In our model only banks will leverage and they do so through bank deposits and repos. Each agent will have a leverage limit which they can not surpass which is sampled from a Pareto distribution.

\subsection{Market Mechanism}

The market works with each agent placing orders and managing their metrics at each time step, during the specified time of the simulation. Agents will have strategies and must manage risk. 

The focus of our analysis is on the default of banks. If a bank can't pay an obligation due to shortage of funds, it becomes defaulted and will enter fire sales to acquire funding. If they lack assets to pay all their obligations or if illiquidity prevents them from turning assets into cash they must declare bankruptcy and are removed from the simulation.

\noindent \textbf{Trading}

At each turn  $t$ each  agent $i$ makes a decision $\epsilon_{i}(t) \in \{-1, 0 , +1\}$, for each asset, where +1 represents the decision to buy, -1 the decision sell and 0 the decision to hold an asset. The decision of each agent is determined by the trader class for stocks, and by it's liquidity limit, VaR limit and leverage limit for the stock fixed-income risky asset and lending operation, and is unrestricted for risk-free assets. Depending on the value of $\epsilon_i(t)$, the trader buys/sells a quantity $q_i(t)$ proportional to his current asset belongings and free cash.

After the decisions are taken the orders are added to the order book \footnote{for simplicity, all orders are placed as market orders. Also, for simplicity, we consider that all the trading is carried out at a centralized desk.}. There is a order book for each asset.

The only price directly affected by orders is the stock price. The way in which order imbalance affects stock prices is similar to the one in \citet{giardina2003bubbles}, where the normalized total order imbalance is used to determine the change of price and is denoted $Q_j(t)$:

\begin{align}
	Q_j(t) = \dfrac{1}{\phi_{j}} \sum_{i=0}^{N} q_{ij}(t) = Q^{+}(t) - Q^{-}(t)
\end{align}

where $\phi_{j}$ is the total number of outstanding shares from the asset(assumed constant in the period of the simulation), $Q^{+}$ is the volume of buy orders and $Q^{-}$ is the volume of sell orders. Trades are cleared and settled in a simple first in first out manner.

\noindent \textbf{Interbank lending}

Banks will use repos to leverage and buy the Risky Asset if their risk limits allow and there is funding available in the money market. Repos are initialized at the market creation and the operations are renewed as long as the banks are bellow their limits and have resources available to renew the operations. At the end of each turn, the banks that need short term liquidity to net their positions will resort to the interbank landing market. 

This sequence of steps is repeated  for the  duration of the simulation and the only changing variable is the trading behavior of agents facing defaults.

\subsection{Failures, Fire sales and Contagion}

Financial institutions may fail under a series of conditions. First, if their assets value fall bellow a threshold they may become insolvent. The same may happen if the agent is run and they are holding illiquid assets.

Failing institutions will enter fire sales, meaning that they will try to liquidate their assets to make cash and terminate their obligations until they become solvent again.  

Fire sales won't affect Government bonds, but will pressure stock and the risky asset prices. The effect in stock prices is felt directly in the price adjustment mechanism, and for the risky asset, it will suffer a discount $d$ for the volume of agents fire selling it. 

Fire sales may also trigger margin calls in repos backed up by the risky asset and will prompt mark-to-market losses. Mark-to-market losses will require an immediate margin call to match the value of the debt and, as a greater number of institutions starts to fire sell the risky asset, it may become more illiquid, requiring an increase in haircut.

Institutions holding a greater amount of illiquid assets will also face higher spreads while trying to acquire funding through interbank loans, expanding their losses. If an agent can't liquidate it's positions and can't obtain funding in the money market it is declared bankrupt and is removed from the simulation.


\subsubsection{Crisis simulation}

Here we describe the structure of the simulation. First, we assume there is a bubble associated with the risky asset which will burst at the beginning of the simulation. As the price of the risky asset loses significant value and liquidity, repos backed up by this asset will require margin calls and higher haircut. These new requirements will make some institutions lose value and deleverage, the institutions unable to meet the new requirements will become defaulted and enter fire sales, triggering new losses in value and liquidity. If they become too illiquid or their losses are substantial, they will not be able to obtain funding in the interbank lending market and must declare bankruptcy, being removed from the simulation. The contagion process is characterized by the following algorithm:

\begin{enumerate}
	\item Initially a bubble collapses, reducing the market value of the risky asset to a $p$ fraction of its original value, $p \in [0,1]$ and causing a reduction in the probability of selling the asset by a fraction $q$, $q \in [0,1]$; 
    
    \item When the market deteriorates, each bank $i$ owning the a $m$ fraction of the risky asset must book mark-to-market losses experiencing a reduction in value by $m * (1 - p)$ and, for each repo operation $j$ of amount $n$ collateralized by the risky asset, there will be a margin call of $(1 - p) * n$  and a haircut increase by the fraction $q$;
    
    \item The loss in value and liquidity will also make a number of agents $k$ be above it's risk limits, triggering a fly to liquidity. These agents will change their trading behavior to sell risky and illiquid assets to obtain more liquid ones. These agents will not fire sell, but will fuel the selling side of order book.
    
    \item When a bank can't meet the new requirements for haircuts they must deleverage, possibly selling it's assets in the process. If the bank faces high illiquidity, it may not be able to deleverage and becomes insolvent, entering fire sales to meet it's obligations.
    
    \item New fire sales will cause new price reductions and lower liquidity, feeding the negative feedback loop.
    
\end{enumerate}

Depending on the simulated scenario the contagion may stop briefly, with no insolvency, or produce a cascade of failures. The model is sensitive to a high number of parameters, such as the asset composition of of the agents, risk limits, distribution of trading behaviors and and severity of the initial crisis. In the next section we perform a simulation to illustrate the model.

\section{Simulation} \label{sec:simulation}

\subsection{Experimental Design}

Our market will be composed of 100 banks, 200 hedge funds and 200 money market funds. Initial NAVs are drawn from a Pareto distribution with parameters $a = 3$ and $m_{banks} = 1e8$, $m_{mmf} = 2e8$ \footnote{Money market funds are created with comparatively high NAVs in order to represent a range of institutional investors and financial intermediaries operating in the money markets.} and $m_{hedge} = 1e7$. While MMFs and Hedge Funds won't start with an initial leverage, banks start with deposits $x \times NAV_bank$ where $x$ follows a Pareto distribution with parameters $a = 3$ and $m = 1$.

\begin{figure}
	\centering
    \includegraphics[width=1\textwidth]{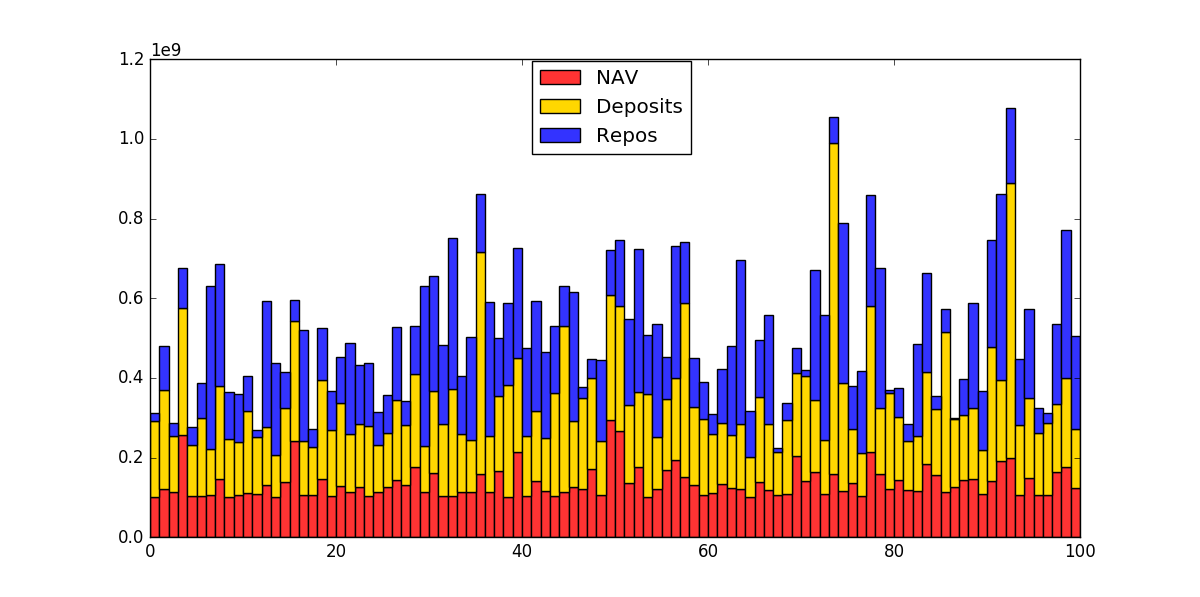}
    \caption{Liability composition of banks in the simulation. After initial market creation liabilities are composed by NAV, bank deposits and repurchase agreements.}
    \label{fig:leverage}
\end{figure}

\begin{figure}
	\centering
    \includegraphics[width=1\textwidth]{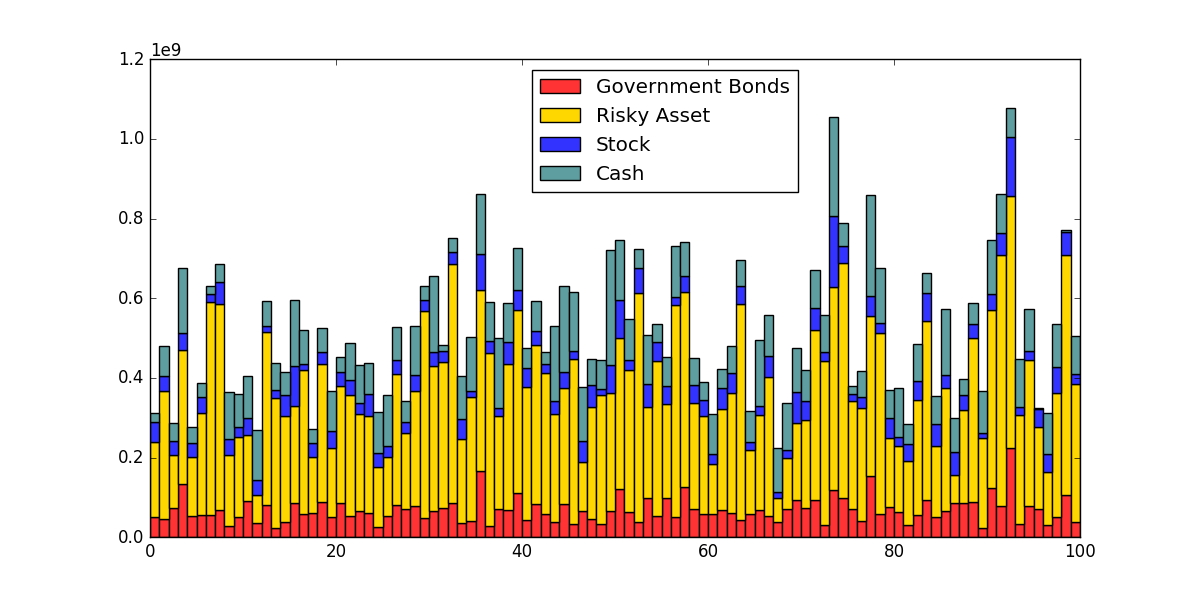}
    \caption{Portfolio composition of banks in the simulation.Banks leverage to invest in the risky asset.}
    \label{fig:portfolio}
\end{figure}

Each bank is assigned a maximum leverage from an uniform distribution with parameters $a = 3$ and $b = 7$. After this assignment there is a random assignment of repo operations between banks and money market funds. Banks use these repo operations to leverage and invest in the Risky Asset. Figure \ref{fig:leverage} shows the composition in the liabilities of the banks and Figure \ref{fig:portfolio} shows their portfolio composition. Other market parameters are defined in Table \ref{tab:parameters}.

\begin{table}[ht]
\centering
\caption{Summary of Financial Agents.}
\label{tab:centrality}
\begin{tabular}{l | l| l | l}
	\hline
	 Agent & Number of Agents &Avg. NAV & Avg. Leverage \\ \hline
	 Banks & 100 & 1.66e8 & 3.781 \\
	 Hedge Funds & 200 & 1.71e7 & 1.000 \\
	 MMFs & 100 & 3.11e8 & 1.000\\
	\hline
\end{tabular}
\end{table}
\bigskip

\begin{table}[ht]
\centering
\caption{Parameters of the initial market.}
\label{tab:parameters}
\begin{tabular}{| l | l| l |}
	\hline
	 Parameter & Symbol & Value \\ \hline
	 Noise traders probabilities & ($p_{b}, p_{s}, p_{h}$) & (0.4, 0.4, 0.2)\\
	 Fundamental traders threshold & $\tau$ & $\mathrm{unif}(.1, .5)$  \\
	 Confidence level for market risk & $\sigma$ & 5\%  \\
     Time horizon of selling probability calculation & n & 3\\
     Leverage limit for banks & $\mathrm{max\_leverage}$ & $\mathrm{unif}(4, 7)$ \\
     
	\hline
\end{tabular}
\end{table}
\bigskip

After the market creation we introduce a shock reducing the liquidity of the Risky Asset by 30\% and its price by 20\%. The results from this simulation are presented next.
\subsection{Results}

\begin{figure}
	\centering
    \includegraphics[width=1\textwidth]{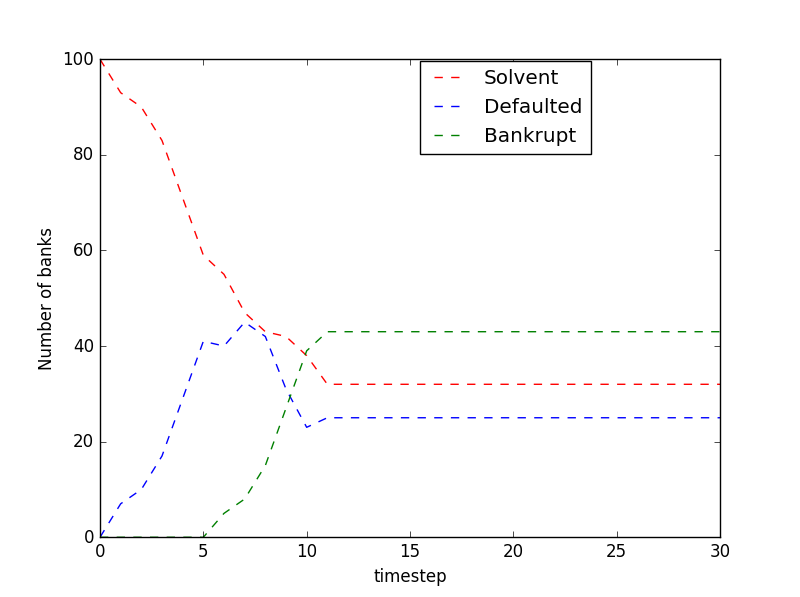}
    \caption{Number of solvent, defaulted and bankrupt at each time step in our simulation. The system reaches a new equilibrium with close to 42\% of the banks having gone bankrupt. }
    \label{fig:failures}
\end{figure}

We focus on the analysis of the crisis over the banking system of our model (i.e. the bank agents). At the beginning of our simulation we have 100 banks, all solvent. Immediately after the shock seven banks become defaulted and the situation worsens until reaching a new equilibrium with 43 banks having gone bankrupt, as is shown in Figure \ref{fig:failures}.

The system also faces massive deleverage with the average leverage falling from 3.781 to 2.644 and the required haircut increases up to 100\%. This means essentially that the repo market has frozen. All the remaining leverage in the results are from bank deposits. In our results, the repo market has evaporated as the result of our shock.

\begin{figure}
	\centering
    \includegraphics[width=1\textwidth]{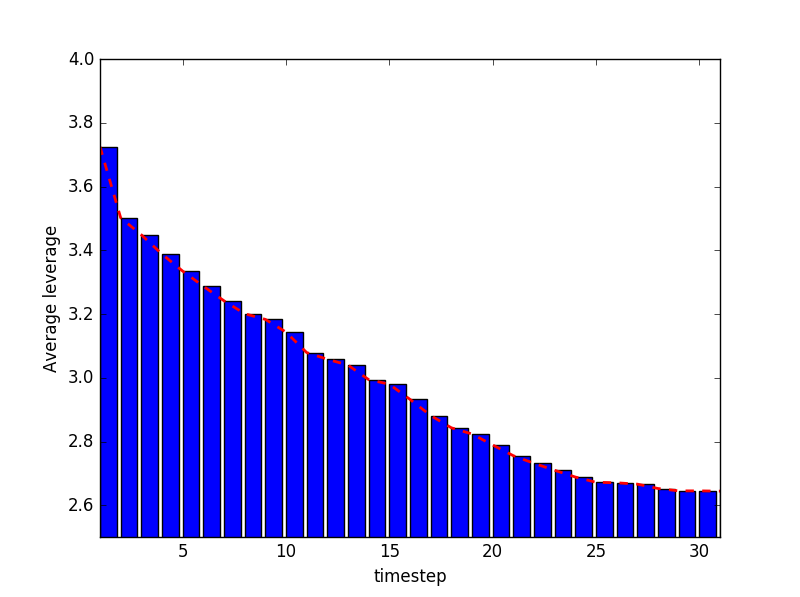}
    \caption{Deleverage of the banking system as the the system suffers a run on the repo.}
    \label{fig:deleverage}
\end{figure}



\begin{figure}
	\centering
    \includegraphics[width=1\textwidth]{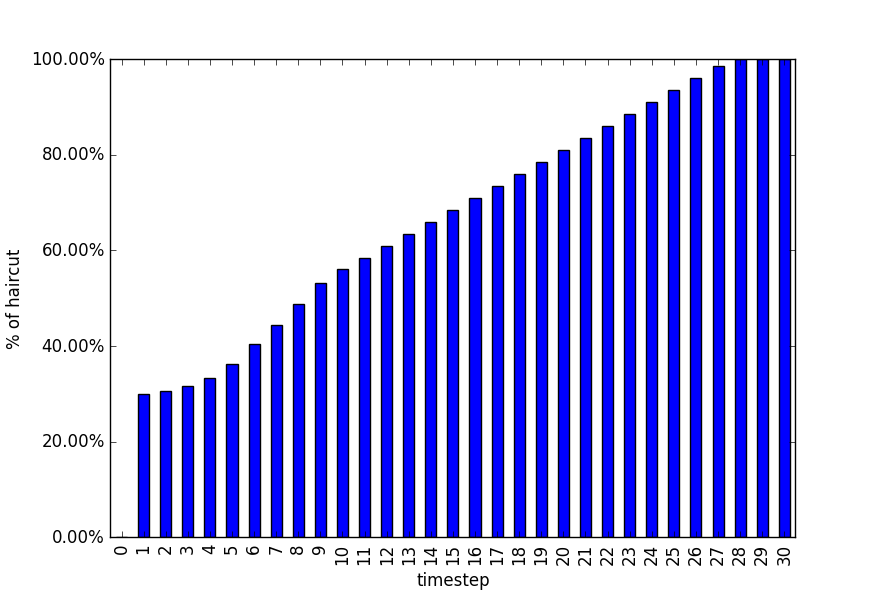}
    \caption{Market required haircut as the financial crisis spreads.}
    \label{fig:haircut}
\end{figure}

\section{Final Remarks} \label{sec:conclusion}

Financial contagion is a complex phenomenon with devastating consequences over the financial system. Here, we have developed a agent-based model of the financial system that recreates some of the observed behaviors of financial institutions in the repo market in order to simulate a crisis propagation. 

While simple, the approach we have developed can be a valuable tool for financial supervisors and financial intermediaries, as they can help assess the stability of the banking system and the resilience of individual participants. For instance, the model can be used with as a stress testing tool to understand the impact of a possible financial crisis over a market segment to guide supervision and investment decisions. 

While the results observed in our simulation are interesting, several improvements are possible to make the model closer to real markets. To advance this line of research: (1) The strategies in the fixed-income market must be modeled to better represent the risk and behavior of participants in this market(2) The effect of bankruptcy over asset prices, especially fixed-income assets, needs to be better determined, and (3)  More assets need to be included in the model, to create a better representation of portfolio allocations.


\bibliographystyle{apalike}

\bibliography{contagion}


\end{document}